\documentclass[twoside,twocolumn,9pt]{article}
\usepackage{extsizes}
\usepackage[super,sort&compress,comma]{natbib} 
\usepackage[version=3]{mhchem}
\usepackage[left=1.5cm, right=1.5cm, top=1.785cm, bottom=2.0cm]{geometry}
\usepackage{balance}
\usepackage{mathptmx}
\usepackage{sectsty}
\usepackage{graphicx} 
\usepackage{lastpage}
\usepackage[format=plain,justification=justified,singlelinecheck=false,font={stretch=1.125,small,sf},labelfont=bf,labelsep=space]{caption}
\usepackage{float}
\usepackage{fancyhdr}
\usepackage{fnpos}
\usepackage[english]{babel}
\addto{\captionsenglish}{%
  \renewcommand{\refname}{Notes and references}
}
\usepackage{array}
\usepackage{droidsans}
\usepackage{charter}
\usepackage[T1]{fontenc}
\usepackage[usenames,dvipsnames]{xcolor}
\usepackage{setspace}
\usepackage[compact]{titlesec}
\usepackage{hyperref}
\definecolor{cream}{RGB}{222,217,201}

\usepackage[export]{adjustbox}
\usepackage{enumitem}
\DeclareUnicodeCharacter{202F}{-}

\usepackage{xr}
\makeatletter

\newcommand*{\addFileDependency}[1]{% argument=file name and extension
\typeout{(#1)}% latexmk will find this if $recorder=0
\@addtofilelist{#1}
\IfFileExists{#1}{}{\typeout{No file #1.}}
}\makeatother

\newcommand*{\myexternaldocument}[1]{%
\externaldocument{#1}%
\addFileDependency{#1.tex}%
\addFileDependency{#1.aux}%
}
\myexternaldocument{ESI}

\begin{document}

\pagestyle{fancy}
\thispagestyle{plain}
\fancypagestyle{plain}{
%%%HEADER%%%
\renewcommand{\headrulewidth}{0pt}
}
%%%END OF HEADER%%%

%%%PAGE SETUP - Please do not change any commands within this section%%%
\makeFNbottom
\makeatletter
\renewcommand\LARGE{\@setfontsize\LARGE{15pt}{17}}
\renewcommand\Large{\@setfontsize\Large{12pt}{14}}
\renewcommand\large{\@setfontsize\large{10pt}{12}}
\renewcommand\footnotesize{\@setfontsize\footnotesize{7pt}{10}}
\makeatother

\renewcommand{\thefootnote}{\fnsymbol{footnote}}
\renewcommand\footnoterule{\vspace*{1pt}% 
\color{cream}\hrule width 3.5in height 0.4pt \color{black}\vspace*{5pt}} 
\setcounter{secnumdepth}{5}

\makeatletter 
\renewcommand\@biblabel[1]{#1}            
\renewcommand\@makefntext[1]% 
{\noindent\makebox[0pt][r]{\@thefnmark\,}#1}
\makeatother 
\renewcommand{\figurename}{\small{Fig.}~}
\sectionfont{\sffamily\Large}
\subsectionfont{\normalsize}
\subsubsectionfont{\bf}
\setstretch{1.125} %In particular, please do not alter this line.
\setlength{\skip\footins}{0.8cm}
\setlength{\footnotesep}{0.25cm}
\setlength{\jot}{10pt}
\titlespacing*{\section}{0pt}{4pt}{4pt}
\titlespacing*{\subsection}{0pt}{15pt}{1pt}
%%%END OF PAGE SETUP%%%

%%%FOOTER%%%
% \fancyfoot{}
% \fancyfoot[LO,RE]{\vspace{-7.1pt}\includegraphics[height=9pt]{head_foot/LF}}
% \fancyfoot[CO]{\vspace{-7.1pt}\hspace{11.9cm}\includegraphics{head_foot/RF}}
% \fancyfoot[CE]{\vspace{-7.2pt}\hspace{-13.2cm}\includegraphics{head_foot/RF}}
% \fancyfoot[RO]{\footnotesize{\sffamily{1--\pageref{LastPage} ~\textbar  \hspace{2pt}\thepage}}}
% \fancyfoot[LE]{\footnotesize{\sffamily{\thepage~\textbar\hspace{4.65cm} 1--\pageref{LastPage}}}}
% \fancyhead{}
% \renewcommand{\headrulewidth}{0pt} 
% \renewcommand{\footrulewidth}{0pt}
% \setlength{\arrayrulewidth}{1pt}
% \setlength{\columnsep}{6.5mm}
% \setlength\bibsep{1pt}
%%%END OF FOOTER%%%

%%%FIGURE SETUP - please do not change any commands within this section%%%
\makeatletter 
\newlength{\figrulesep} 
\setlength{\figrulesep}{0.5\textfloatsep} 

\newcommand{\topfigrule}{\vspace*{-1pt}% 
\noindent{\color{cream}\rule[-\figrulesep]{\columnwidth}{1.5pt}} }

\newcommand{\botfigrule}{\vspace*{-2pt}% 
\noindent{\color{cream}\rule[\figrulesep]{\columnwidth}{1.5pt}} }

\newcommand{\dblfigrule}{\vspace*{-1pt}% 
\noindent{\color{cream}\rule[-\figrulesep]{\textwidth}{1.5pt}} }

\makeatother
%%%END OF FIGURE SETUP%%%

%%%TITLE, AUTHORS AND ABSTRACT%%%
\twocolumn[
  \begin{@twocolumnfalse}
% {\includegraphics[height=30pt]{head_foot/PCCP}\hfill\raisebox{0pt}[0pt][0pt]{\includegraphics[height=55pt]{head_foot/RSC_LOGO_CMYK}}\\[1ex]
% \includegraphics[width=18.5cm]{head_foot/header_bar}}\par
\vspace{1em}
\sffamily
\begin{tabular}{m{4.5cm} p{13.5cm} }

 & \noindent\LARGE{\textbf{Relaxation enhancement by microwave irradiation may limit dynamic nuclear polarization}} \\%Article title goes here instead of the text "This is the title"
\vspace{0.5cm} & \vspace{0.5cm} \\

 & \noindent\large{Gevin von Witte,\textit{$^{a,b}$} Aaron Himmler,\textit{$^{b}$} Sebastian Kozerke,\textit{$^{a}$} and Matthias Ernst\textit{$^{b \ast}$}} \\%Author names go here instead of "Full name", etc.

\vspace{0.5cm} & \vspace{0.5cm} \\

& \noindent\normalsize{Dynamic nuclear polarization enables the hyperpolarization of nuclear spins  beyond the thermal-equilibrium Boltzmann distribution.
However, it is often unclear why the experimentally measured hyperpolarization is below the theoretical achievable maximum polarization. 
We report a (near-) resonant relaxation enhancement by microwave (MW) irradiation, leading to a significant increase in the nuclear polarization decay compared to measurements without MW irradiation. For example, the increased nuclear relaxation limits the achievable polarization levels to around 35\% instead of hypothetical 60\%, measured in the  DNP material TEMPO in \textsuperscript{1}H glassy matrices at 3.3\;K and 7\;T. 
Applying rate-equation models to published build-up and decay data indicates that such relaxation enhancement is a common issue in many samples when using different radicals at low sample temperatures and high Boltzmann polarizations of the electrons. 
Accordingly, quantification and a better understanding of the relaxation processes under MW irradiation might help to design samples and processes towards achieving higher nuclear hyperpolarization levels.} \\%The abstrast goes here instead of the text "The abstract should be..."

\end{tabular}

 \end{@twocolumnfalse} \vspace{0.6cm}

  ]
%%%END OF TITLE, AUTHORS AND ABSTRACT%%%

%%%FONT SETUP - please do not change any commands within this section
\renewcommand*\rmdefault{bch}\normalfont\upshape
\rmfamily
\section*{}
\vspace{-1cm}

%%%FOOTNOTES%%%

\footnotetext{\textit{$^{a}$~Institute for Biomedical Engineering, University and ETH Zurich, 8092 Zurich, Switzerland}}
\footnotetext{\textit{$^{b}$~Department of Chemistry and Applied Biosciences, ETH Zurich, 8093 Zurich, Switzerland }}
\footnotetext{\textit{$^{\ast}$~Corresponding author: maer@ethz.ch}}

%Please use \dag to cite the ESI in the main text of the article.
%If you article does not have ESI please remove the the \dag symbol from the title and the footnotetext below.
% \footnotetext{\dag~Electronic Supplementary Information (ESI) available: [details of any supplementary information available should be included here]. See DOI: 10.1039/cXCP00000x/}
%additional addresses can be cited as above using the lower-case letters, c, d, e... If all authors are from the same address, no letter is required

%\footnotetext{\ddag~Additional footnotes to the title and authors can be included \textit{e.g.}\ `Present address:' or `These authors contributed equally to this work' as above using the symbols: \ddag, \textsection, and \P. Please place the appropriate symbol next to the author's name and include a \texttt{\textbackslash footnotetext} entry in the the correct place in the list.}

%%%END OF FOOTNOTES%%%

%%%MAIN TEXT%%%%
\section*{Introduction} % The \section*{} command stops section numbering
	
%\addcontentsline{toc}{section}{Introduction} % Adds this section to the table of contents

Methods to generate hyperpolarization on nuclei have a broad field of application in nuclear magnetic resonance (NMR) \cite{ardenkjaer-larsen_facing_2015,Thankamony2017}, magnetic-resonance imaging (MRI) \cite{Nelson2013,Mugler2013,Wang2019a,Gallagher2020,Fuetterer2022}, particle physics \cite{Crabb1997,Kowalska2021} and quantum science \cite{Ajoy2019,Bucher2020,Alvarez2015,Broadway2018,McCamey2009,Gangloff2019}. 
Hyperpolarization can be generated by many different methods \cite{Eills2023} although reaching the theoretical maximum for a given method remains experimentally challenging. 
The polarization-transfer process is a poorly understood problem and it is difficult to assess what actually limits the enhancements. 
In many cases, hyperpolarization is generated by a small but continuous injection of polarization which, in the absence of relaxation, should enable polarization levels up to the theoretical limit.

Dynamic nuclear polarization (DNP) is a hyperpolarization method that is based on the transfer of polarization (spin magnetic moment) from unbound valence electrons (radicals, defects, paramagnetic centers) to hyperfine-coupled nuclear spins under microwave (MW) irradiation. 
DNP experiments at cryogenic temperatures offer the advantages of long electron relaxation time and increased Boltzmann polarization of the electrons to near unity.
Typically, the electron spins are very dilute in large nuclear spins systems with on average hundreds of nuclear spins per electron. 
Such a configuration leads to a slow build up of hyperpolarization as successively more and more nuclear spins are polarized either directly from the electrons or indirectly through nuclear spin diffusion. 
Eventually a steady-state polarization ($P_0$) is reached where nuclear relaxation and polarization injection compensate each other. 
The build-up of the polarization can often be described by a characteristic time constant $\tau_\mathrm{bup}$. 
Upon switching off the MW irradiation, the hyperpolarized state relaxes back to the thermal equilibrium, commonly called polarization decay and described by a characteristic time constant $\tau_\mathrm{decay}$. 
%The nuclear spin-lattice relaxation time $T_{1,n}$ in combination with spin diffusion will determine the decay constant. 
Rate equation or kinetic models offer the possibility to qualitatively and even quantitatively describe the balance between hyperpolarization injection and relaxation \cite{von_witte_modelling_2023}.

In this paper we use such rate equation models to determine the relaxation-rate constant of the nuclei during the DNP polarization build-up process. 
We find that in many cases the nuclear relaxation-rate constants during the hyperpolarization build-up under MW irradiation are much larger than the ones obtained without MW irradiation. 
Microwave irradiation was previously shown to shorten rotating-frame relaxation times ($T_{1\rho}$) in cross-polarization experiments \cite{Bornet2016} and also nuclear coherence life times ($T_2'$) \cite{Guarin2022} at liquid helium temperatures and 6.7\,T field strength.
In this work, the increased nuclear relaxation-rate constants under microwave irradiation during the build up of hyperpolarization is found to limit the achievable maximum polarization compared to the prediction when using the electron equilibrium polarization, nuclear relaxation-rate constants (without MW irradiation) and polarization injection rates.

\section*{Methods}

\subsection*{Samples}

All measurements were performed with 50\;mM 4-oxo-TEMPO in water/ glycerol mixtures and DNP performed on the \textsuperscript{1}H spins. 
In particular, we compare two different sample formulations with (i) TEMPO in a (6:3:1)\textsubscript{V} mixture of glycerol-d\textsubscript{8}, D\textsubscript{2}O and H\textsubscript{2}O (DNP juice) and (ii) TEMPO in (1/1)\textsubscript{V} H\textsubscript{2}O/ glycerol (no deuteration, all natural abundance).
After mixing the ingredients, the filled sample container was frozen in liquid nitrogen before being transferred to the cryogenically pre-cooled polarizer (cryostat temperature during the transfer below 20K). 
All chemicals were purchased commercially and used without further purification.
To check the reproducibility of the results, we tested two independently formulated natural abundance samples.  
The measurements were performed together with those reported in Ref. \cite{von_witte_modelling_2023}.

\subsection*{DNP and EPR}

The DNP measurements were performed on a home-built polarizer \cite{Jahnig2017} at a temperature of 3.3\;K unless otherwise indicated. The system was equipped with a Bruker Avance III (Bruker BioSpin AG, Switzerland) spectrometer at 7\;T  (299\;MHz \textsuperscript{1}H Larmor frequency).
The natural-abundance sample formulation was reported to build up mono-exponentially under the same conditions before \cite{Jahnig2019}. 
Compared to the previously published work on this material, the polarizer was upgraded to a more powerful MW source (200\;mW, Virginia Diodes Inc., USA) and in-house electroplated low-loss wave guides, giving us approximately eight times more MW power as before at the sample space (around 65\;mW at the sample) \cite{Himmler2022}. 
The other details of the set-up are described elsewhere and were unchanged \cite{Jahnig2017,Himmler2022}. 
The absolute values of the polarization were calculated on a single thermal equilibrium measurement per sample as reported and discussed previously \cite{von_witte_modelling_2023}.

Electron paramagnetic resonance (EPR) spectra were acquired with our in-house developed longitudinal-detection (LOD) EPR set-up at 7\;T at a temperature of 5\;K \cite{Himmler2022}.  

For all build-up curves (except where otherwise stated to check the off-resonant irradiation), the MW frequency was set to 197.05\;GHz without using any MW frequency modulation. 
The MW power was set to the maximum output unless explicitly stated. 
For decay curves under MW irradiation, the MW frequency was switched to 197.28\;GHz (zero crossing of the DNP profile) after the sample (nearly) reached its steady-state polarization using maximum output power unless otherwise stated.

All data processing was performed with in-house developed MATLAB (MathWorks Inc., Natick, Massachusetts, U.S.A) scripts. 
All experimental uncertainties in the processed experimental data result from the 95\% fit intervals.
Experimental instabilities like slight changes in the MW output or minor temperature fluctuations as well as uncertainties in the thermal equilibrium measurement were not included in the analysis. 

\section*{Results and discussion}
\subsection*{Hyperpolarization decays under microwave irradiation}

\begin{figure*}[!ht]\centering
	\includegraphics[width=\linewidth]{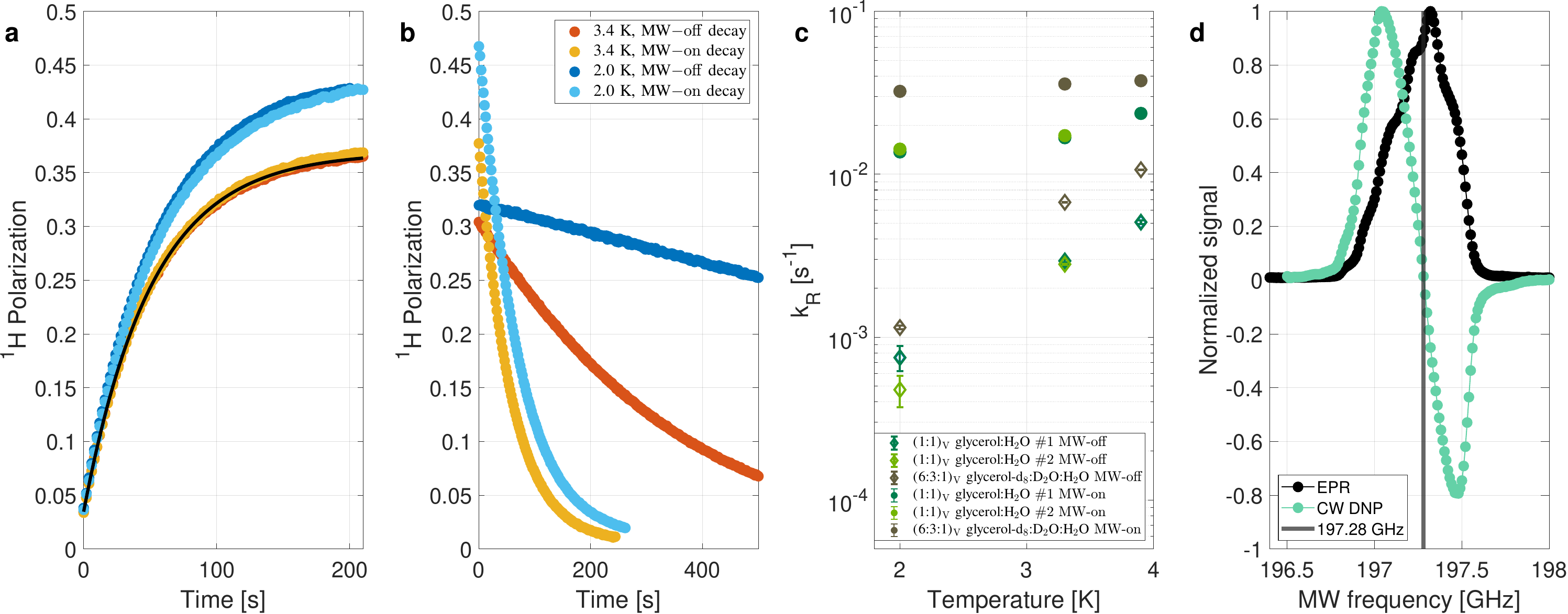}
	\caption{Nuclear relaxation depending on the temperature and MW irradiation. Build-ups \textbf{(a)} and the respective decays \textbf{(b)} of the (second) natural abundance sample at 3.3\;K (red and orange) and 2.0\;K (light and darker blue) without correcting for the effects of radio-frequency (RF) pulses. The MW is set to the same power and frequency (197.05\;GHz) during all build-ups. During the decays depicted in darker blue and red, the MW is switched off. For the decays depcited in orange (3.3\;K) and light blue (2.0\;K), the MW is switched to the central EPR frequency (197.28\;GHz) resulting in zero DNP, as can be seen in (d). The measured polarization in (a) can be simulated (solid black line) with a single compartment rate equation model \cite{von_witte_modelling_2023}. The only model input parameters are the thermal electron polarization, measured build-up time and steady-state polarization. \textbf{(c)} Relaxation rates during MW-on decays (dots) and MW-off decays (diamonds) for the natural abundance samples (green) and partially deuterated (DNP juice) sample (brown-gray/taupe, see Methods for details). The data shown are without RF correction (explained in the main text) \textbf{(d)} Overlay of the EPR line at 5\;K (blue) from \cite{Himmler2022} and a DNP profile. For simplicity, all measurements presented in this work were recorded with continuous wave (CW) irradiation without MW modulation.}
	\label{fig:BupDecayOverview}
\end{figure*}

Figure \ref{fig:BupDecayOverview} shows polarization build-up (a) and decay (b) curves for the natural-abundance sample at two different temperatures, 3.3\;K (red and yellow lines) and 2.0\;K (blue lines) corresponding to 89 and 99\% thermal electron polarization at 7\;T.
The two sets of build-up curves (Fig.~\ref{fig:BupDecayOverview}a) were recorded under identical conditions and show the good reproducibility of the measurements.
Due to the higher thermal polarization at 2\;K, the build-up curves at lower temperature (blue curves) reach a higher steady-state polarization.

The corresponding decays shown in Fig.~\ref{fig:BupDecayOverview}b were recorded with the MW source switched off (dark blue and red curves) as well as with the MW source turned on (light blue and yellow curves) but tuned to the frequency of the zero crossing of the DNP profile (197.28\;GHz, see Fig.~\ref{fig:BupDecayOverview}d), minimizing the DNP enhancement during the decay measurements. 
There are several effects visible when comparing the decay curves with and without MW irradiation: 
\begin{enumerate}[wide, label=(\roman*), labelwidth=!, labelindent=0pt]
\item Without MW irradiation, the decay of the nuclear polarization is slower (dark blue and red line). 
Switching the MW irradiation to the frequency of the zero crossing of the DNP profile significantly speeds up the nuclear spin-lattice ($T_1$) relaxation (light blue and yellow line in Fig.~\ref{fig:BupDecayOverview}b). 
\item The decay curves with on-resonance MW irradiation (light blue and yellow lines in Fig.~\ref{fig:BupDecayOverview}b) start at a higher value, the ones without MW irradiation (dark blue and red lines in Fig.~\ref{fig:BupDecayOverview}b) at a lower value than the end point of the build-up curves.  This can be explained by the presence of strongly hyperfine-coupled and, therefore, invisible (hidden or quenched) nuclear spins. Microwave irradiation will decouple some of these spins and make them visible and contribute to the observable nuclear signal intensity \cite{Saliba2017,Tan2022}. 
\item In the low temperature (2.0\;K), MW-off decays corrected for the perturbations by RF pulses shown in Fig.~\ref{fig:BupDecayOverview_corr} (dark blue), the observed signal first increases slightly, in contrast to the expected decay as observed in the uncorrected data (Fig.~\ref{fig:BupDecayOverview}b and Fig.~\ref{fig:BupDecayOverview_uncorr}). This can again be explained by invisible strongly hyperfine-shifted spins that are coupled to the visible spins by slow spin diffusion\cite{Stern2021}.
\end{enumerate}

In the following, we will not discuss the latter two points but focus on the first point. 
Such a dependence of nuclear decay rates on the MW irradiation at low temperatures has not yet been reported to the best of our knowledge. 
Similar and related effects, i.e., a decrease of rotating-frame relaxation times ($T_{1\rho}$) in cross-polarization experiments under MW irradiation \cite{Bornet2016} and a decrease of transverse nuclear coherence life times ($T_2'$) in echo experiments \cite{Guarin2022} with MW irradiation at liquid-He temperatures have been reported recently.
 
Fig.~\ref{fig:BupDecayOverview}c shows the fitted decay-rate constants for three different samples with and without MW irradiation at three different temperatures. 
In all cases, a significant increase of the nuclear decay-rate constant is observed under MW irradiation. 
Under MW irradiation, the relaxation is nearly independent of the temperature between 2.0 and 3.9\;K, contrasting the nearly one order of magnitude difference in decay-rate constants without MW irradiation over this range of temperatures.

\begin{figure*}[!ht]\centering % Using \begin{figure*} makes the figure take up the entire width of the page
	\includegraphics[max width=\textwidth]{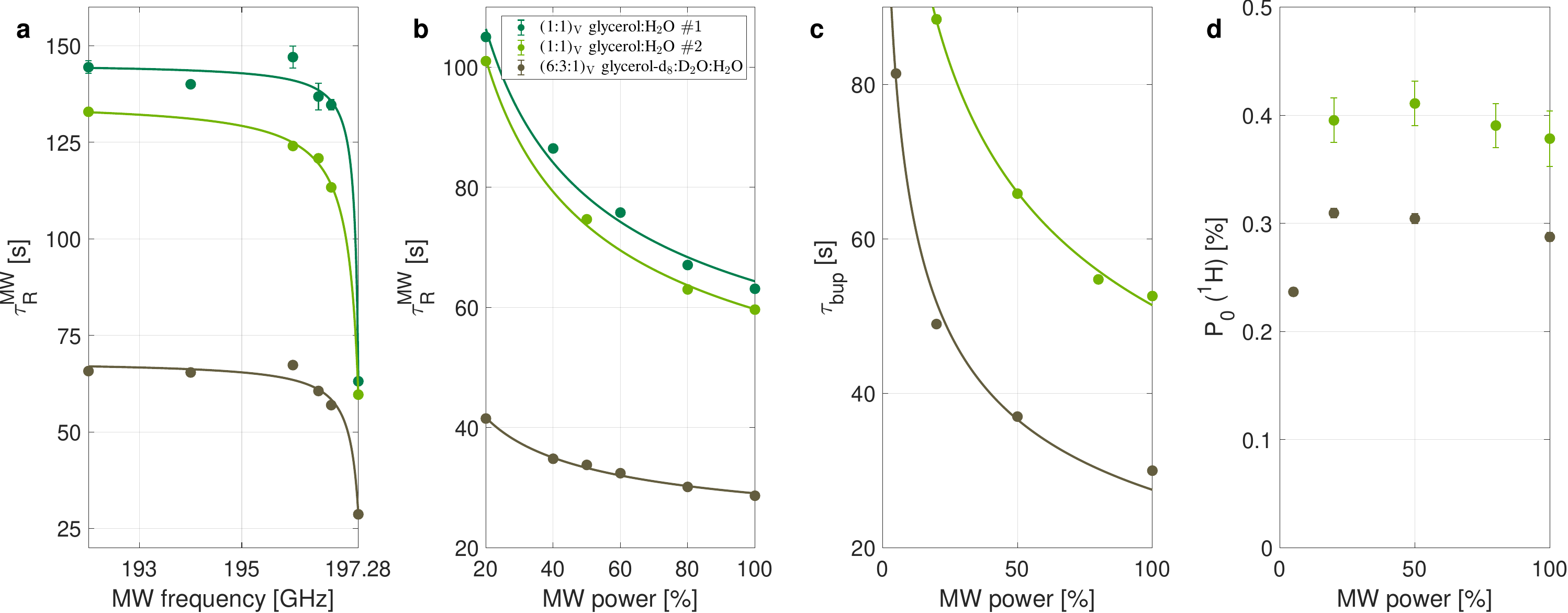}
	\caption{Microwave-dependence of the time constants and polarization. Frequency \textbf{(a)} and power dependence \textbf{(b)} of the relaxation enhancement during MW irradiation in decay experiments. For the frequency dependence, the MW was set to maximum power. All power dependencies were recorded at the central zero-DNP frequency (197.28\;GHz). The data in (a) is fitted with $\tau_\mathrm{R}^{\nu_\mathrm{MW}=0}-m/(c+|197.28-{\nu_\mathrm{MW}}|)$. Build-up time \textbf{(c)} and steady-state polarization \textbf{(d)} show the same power-dependence for natural abundance (green) and partial deuteration (brown-gray/ taupe). The steady-state polarization is relatively independent of the MW-power except for small powers. Data in (b) and (c) were both fitted with a square-root ansatz ($\propto P_\mathrm{MW}^{-1/2}$ with the microwave power $P_\mathrm{MW}$).}
	\label{fig:Experiment_1compartment_Relaxation_Microwave}
\end{figure*}

To investigate this further, we measured the decay-rate constants as a function of the MW settings (power and frequency) which is summarized in Fig. \ref{fig:Experiment_1compartment_Relaxation_Microwave}a and b. 
The relaxation enhancement shows a frequency-dependent resonant behavior with stronger relaxation enhancement close to the zero crossing of the DNP profile which corresponds to the center of the EPR spectrum (Fig.~\ref{fig:BupDecayOverview}d).
For a DNP frequency of 197.28\;GHz (zero crossing of the DNP profile), the power dependence of the relaxation enhancement can be described with a square-root scaling (Fig. \ref{fig:Experiment_1compartment_Relaxation_Microwave}b) and scales in the same way as the  build-up time constant (see Fig.~\ref{fig:Experiment_1compartment_Relaxation_Microwave}c) as a function of MW power.
The steady-state polarization $P_0$, however, shows a different behavior as shown in Fig.~\ref{fig:Experiment_1compartment_Relaxation_Microwave}d: It is nearly independent of the power between 20 and 100\% of the MW output. 

To understand the almost MW-power independent steady-state polarization while the decay-rate as well as build-up time constant $\tau_\mathrm{bup}$ follow a square root scaling, we use a single compartment model\cite{von_witte_modelling_2023}.
For a single homogeneous compartment, the hyperpolarization build-up can be described using a first-order differential equation with a hyperpolarization injection rate constant $k_\mathrm{W}$ and a relaxation-rate constant of the build-up $k_\mathrm{R}^\mathrm{bup}$
\begin{align}
	\frac{\text{d}P}{\text{d}t} = (A-P) k_\mathrm{W} - k_\mathrm{R}^\mathrm{bup} P \label{eq:ODE_bup_1compartment}
\end{align}
with $A$ describing the theoretical maximum of hyperpolarization achievable, i.e., the thermal electron polarization in DNP.
The solution of Eq.~\ref{eq:ODE_bup_1compartment} is a mono-exponential curve which can be compared with the phenomenological description of the build-up curve using $P(t) = P_0 (1-e^{-t/\tau_\mathrm{bup}})$ to express the experimental parameters in terms of model parameters. 
Here, $P_0$ is the steady-state polarization and $\tau_{\mathrm{bup}}$ the build-up time.
\begin{subequations} \label{eq:1compartment_bup_solution}
	\begin{align}
		\tau_{\mathrm{bup}}^{-1} &= k_\mathrm{W}+k_\mathrm{R}^\mathrm{bup} \label{eq:1compartment_bup_tau} \\
		P_0 &= \frac{Ak_\mathrm{W}}{k_{W}+k_\mathrm{R}^\mathrm{bup}} = Ak_\mathrm{W}\tau_{\mathrm{bup}} \label{eq:1compartment_bup_P0}
	\end{align}
\end{subequations}
 
For the decay, $k_\mathrm{W}$ would be set to zero (MW off), leading to $\tau_\mathrm{decay}^{-1}=\tau_R^{-1}=k_\mathrm{R}^\mathrm{decay}$. 
The model can reproduce the experimental build-up curve as shown in Fig.~\ref{fig:BupDecayOverview}a (black line). 
A different approach to measure the rate-equation parameters is discussed in Sec.~\ref{sec:SI_ModelValidation} of the supplementary information.

From the definition of the build-up time in Eq.~\ref{eq:1compartment_bup_tau} and the observed square root scaling of $\tau_\mathrm{decay}$ and $\tau_\mathrm{bup}$ with MW power, the injection rate constant, $k_{W}$ will obey a power scaling similar to a square root. 
Since all rates in the steady-state polarization (cf. Eq.~\ref{eq:1compartment_bup_P0}) follow roughly a square root scaling, a power-independent steady-state polarization is expected. 
Therefore, higher MW power will lead to a faster build-up but not to a higher steady-state polarization.

However, this does not hold true for low MW powers as in this case the relaxation enhancement under MW irradiation is small compared to the thermal relaxation and the total relaxation-rate constant will not scale with the square root of the MW power. 
Hence, the total relaxation in the system competing with the DNP injection can be described as $k_\mathrm{R} = k_\mathrm{R}^\mathrm{decay}+k_\mathrm{R}^{MW}sB_{1,MW}$ with the MW-field amplitude $B_{1,MW}\propto \sqrt{P_\mathrm{MW}}$ (MW power $P_\mathrm{MW}$) and a scaling factor $s$. 
For small MW powers, the total relaxation in the system is dominated by thermal relaxation, explaining the increase of the steady-state polarization with MW power as only the injection-rate constant grows with the square root of the MW power (linear in $B_{1,\mathrm{MW}}$). 
For larger MW powers, MW induced relaxation governs the total relaxation and no further improvements of the steady-state polarization and only a faster build-up is expected (see Fig.~\ref{fig:Experiment_1compartment_Relaxation_Microwave}). 
A shortening of the build-up time with a preserved enhancement might help to improve the NMR signal per unit time \cite{Leavesley2018a}. 

Extracting the rate-equation parameters $k_\mathrm{R}$ and $k_\mathrm{W}$ directly from the experimental build-up time constant and steady-state polarization value using Eq.~\ref{eq:1compartment_bup_solution} leads to relaxation-rate constants three to four times larger than the decay-rate constant measured without MW irradiation (compare Figs. \ref{fig:BupDecayOverview}c and \ref{fig:Experiment_1compartment_TheoryParameters}c). 
Using the injection-rate constant from the highest applied MW power and assuming the decay-rate constant measured without MW irradiation, the steady-state polarization would be around 60-70\% instead of the experimentally measured 30-40\% as shown in Fig.~\ref{fig:HypotheticalPolarization}.

The injection rate as used in Eq.~\ref{eq:ODE_bup_1compartment} refers to the creation of polarization (normalized magnetization) and not to absolute magnetization which would describe the absolute number of effective electron-nuclear polarization transfers.
In our experiments. all samples contained the same concentration of radicals (50\;mM). 
The approximately 90\% deuteration reduced the number of protons per radical. 
Hence, the similar (polarization) injection rates for deuterated and natural abundance samples (cf. Fig.~\ref{fig:Experiment_1compartment_TheoryParameters}c) indicate that DNP (at least in the partially deuterated sample) is limited by transport from the strongly hyperfine coupled spins into the bulk rather than by the initial electron-to-nucleus polarization transfer: Similar steady-state polarizations are found for natural abundance and partially deuterated samples (cf. Fig.~\ref{fig:HypotheticalPolarization}) which requires around ten times more nuclear spins to be polarized per TEMPO radical in the the natural abundance sample. If the electron-to-nuclear polarization transfer would be the DNP limiting step, a ten times higher (polarization) injection rate would be expected for the deuterated sample. 

%Adjusting for the different \textsuperscript{1}H nuclei concentrations between the natural abundance and partially deuterated sample, would give an approximately ten times higher number of nuclei polarized per second and radical for the natural abundance sample. 
%If the initial electron-to-nucleus polarization transfer would be rate limiting, the polarization injection for the deuterated sample should be approximately ten times higher than for the natural abundance sample.
Similar conclusions have been made for partially deuterated samples \cite{Prisco2021}, termed "spin-diffusion limited" samples, with a somewhat more complicated model.
For high \textsuperscript{1}H concentrations, Ref. \cite{Prisco2021} ascribes the polarization to be "spin-exchange limited" as too many nuclei per radical need to be polarized, possibly supported by the slight decrease in $k_\mathrm{W}$ for the natural abundance samples compared to the partially deuterated sample (cf. Fig.~\ref{fig:Experiment_1compartment_TheoryParameters}c). 

To investigate whether an enhancement of the apparent decay rates under MW irradiation is a particular property of our experimental sample composition, we have reevaluated data available in the literature using the one-compartment model. 
Using experimental build-up and decay times as well as steady-state polarization levels allows us to calculate and compare the relaxation-rate constants during build up and decay. 

\subsection*{Relaxation enhancement by MW irradiation across samples and DNP conditions}

\begin{table*}\centering % Using \begin{figure*} makes the figure take up the entire width of the page
\caption{Summary of relaxation enhancements \textsuperscript{\emph{a}} by MW irradiation in different samples\textsuperscript{\emph{b}} and conditions.}
\begin{center}
\begin{adjustbox}{width=\textwidth}
\label{Tab:RelaxationEnhancementLiterature}
\begin{tabular}{l c c|c c c|c c c|c c c c r}
 &  & $c(Gd)$ & $B_0$ & T & $c(e^-)$ & $P_0$ & $\tau_\mathrm{bup}$  & $T_1$ & $k_\mathrm{W}$ & $k_\mathrm{R}^\mathrm{bup}$ & $k_\mathrm{R}^\mathrm{decay}$ & $k_\mathrm{R}^\mathrm{bup}/k_\mathrm{R}^\mathrm{decay}$ & Ref. \\ 
 &  & mM & T & K & mM & \% & s & s & $(10^{3} s)^{-1}$ & $(10^{3} s)^{-1}$ & $(10^{3} s)^{-1}$ &  &  \\ \hline
Trityl & \textsuperscript{13}C  &  & 3.35 & 3.5 & 15 & 3.5 & 353 & 419 & 0.17 & 2.66 & 2.39 & 1.1 & \cite{BatelThesis} \\ 
Trityl & \textsuperscript{13}C  &  & 3.35 & 1.0 & 15 & 27 & 1200 & 10000 & 0.23 & 0.60 & 0.10 & 6.0 & \cite{macholl_trityl_2010} \\ 
Trityl & \textsuperscript{13}C  & 1.8 & 3.35 & 1.0 & 14.9 & 50 & 1500 & 9000 & 0.34 & 0.33 & 0.11 & 2.9 & \cite{macholl_trityl_2010} \\
Trityl & \textsuperscript{13}C  &  & 4.64 & 1.0 & 14.3 & 64 & 3000 & 20000 & 0.21 & 0.12 & 0.05 & 2.4 & \cite{macholl_trityl_2010} \\ 
Trityl & \textsuperscript{13}C  &  & 4.64 & 1.0 & 18.5 & 58 & 1600 & 22000 & 0.36 & 0.26 & 0.05 & 5.7 & \cite{macholl_trityl_2010} \\ \hline
Trityl & \textsuperscript{13}C  &  & 4.64 & 1.0 & 45.4 & 23 & 475 & 9000 & 0.49 & 1.62 & 0.11 & 15 & \cite{macholl_trityl_2010} \\ 
Trityl & \textsuperscript{13}C  & 1.5 & 4.64 & 1.0 & 14.3 & 70 & 3000 & 14000 & 0.23 & 0.10 & 0.07 & 1.4 & \cite{macholl_trityl_2010} \\ 
Trityl & \textsuperscript{13}C  &  & 3.35 & 1.3 & 15 & 27 & 1000 & 6400 & 0.29 & 0.71 & 0.16 & 4.6 & \cite{Ardenkjaer-Larsen2019,Ardenkjaer-LarsenPrivate} \\ 
Trityl & \textsuperscript{13}C  &  & 6.7 & 1.3 & 15 & 65 & 3700 & 33000 & 0.18 & 0.09 & 0.03 & 3.1 & \cite{Ardenkjaer-Larsen2019,Ardenkjaer-LarsenPrivate} \\ 
Trityl & \textsuperscript{13}C  &  & 6.7 & 1.3 & 30 & 70 & 1200 & 33000 & 0.58 & 0.25 & 0.03 & 8.2 & \cite{Ardenkjaer-Larsen2019,Ardenkjaer-LarsenPrivate} \\ \hline
Trityl & \textsuperscript{13}C  &  & 10.1 & 1.3 & 15 & 70 & 26000 & 90000 & 0.03 & 0.01 & 0.01 & 1.0 & \cite{Ardenkjaer-Larsen2019,Ardenkjaer-LarsenPrivate} \\ 
Trityl & \textsuperscript{13}C  &  & 10.1 & 1.3 & 45 & 70 & 2200 & 90000 & 0.32 & 0.14 & 0.01 & 12 & \cite{Ardenkjaer-Larsen2019,Ardenkjaer-LarsenPrivate} \\ 
Trityl & \textsuperscript{1}H &  & 7 & 20 & 10 & 0.0 & 80 & 85 & 0.01 & 12.5 & 11.8 & 1.1 & \cite{Equbal2020} \\ 
Trityl & \textsuperscript{1}H &  & 7 & 20 & 40 & 0.3 & 40 & 50 & 0.29 & 24.7 & 20.0 & 1.2 & \cite{Equbal2020} \\ 
Trityl & \textsuperscript{1}H &  & 7 & 20 & 100 & 1.0 & 1.75 & 2.5 & 24.5 & 547 & 400 & 1.4 & \cite{Equbal2020} \\ \hline
TEMPO & \textsuperscript{13}C  &  & 3.35 & 4.2 & 25 & 1.0 & 346 & 420 & 0.06 & 2.83 & 2.38 & 1.2 & \cite{Jahnig2019} \\ 
TEMPO & \textsuperscript{13}C  &  & 3.35 & 4.2 & 50 & 1.4 & 117 & 169 & 0.24 & 8.30 & 5.92 & 1.4 & \cite{Jahnig2019} \\ 
TEMPO & \textsuperscript{13}C  &  & 3.35 & 4.2 & 75 & 1.3 & 57 & 91 & 0.45 & 17.1 & 11.0 & 1.6 & \cite{Jahnig2019} \\ 
TEMPO & \textsuperscript{1}H  &  & 3.35 & 4.2 & 25 & 3.7 & 102 & 192 & 0.75 & 9.05 & 5.21 & 1.7 & \cite{Jahnig2019} \\
TEMPO & \textsuperscript{1}H  &  & 3.35 & 4.2 & 50 & 4.7 & 34 & 84 & 2.84 & 26.6 & 11.9 & 2.2 & \cite{Jahnig2019} \\ \hline
TEMPO & \textsuperscript{1}H  &  & 3.35 & 4.2 & 75 & 4.5 & 18 & 42 & 5.08 & 50.5 & 23.8 & 2.1 & \cite{Jahnig2019} \\ 
TEMPO & \textsuperscript{13}C  &  & 7 & 4.2 & 25 & 3.6 & 1271 & 1850 & 0.04 & 0.75 & 0.54 & 1.4 & \cite{Jahnig2019} \\ 
TEMPO & \textsuperscript{13}C  &  & 7 & 4.2 & 50 & 4.4 & 414 & 666 & 0.13 & 2.28 & 1.50 & 1.5 & \cite{Jahnig2019} \\ 
TEMPO & \textsuperscript{13}C  &  & 7 & 4.2 & 75 & 4.2 & 205 & 349 & 0.26 & 4.62 & 2.87 & 1.6 & \cite{Jahnig2019} \\ 
TEMPO & \textsuperscript{1}H  &  & 7 & 4.2 & 25 & 7.7 & 242 & 439 & 0.39 & 3.74 & 2.28 & 1.6 & \cite{Jahnig2019} \\ \hline
TEMPO & \textsuperscript{1}H  &  & 7 & 4.2 & 50 & 18.4 & 81 & 216 & 2.81 & 9.54 & 4.63 & 2.1 & \cite{Jahnig2019} \\ 
TEMPO & \textsuperscript{1}H  &  & 7 & 4.2 & 75 & 16.7 & 39 & 106 & 5.30 & 20.3 & 9.43 & 2.2 & \cite{Jahnig2019} \\ 
TEMPO & \textsuperscript{1}H  &  & 7 & 3.9 & 50 & 27 & 40 & 197 & 7.99 & 17.0 & 5.08 & 3.4 & This work \\ 
TEMPO & \textsuperscript{1}H  &  & 7 & 3.3 & 50 & 37 & 54 & 340 & 7.63 & 10.9 & 2.94 & 3.7 & This work \\ 
TEMPO & \textsuperscript{1}H  &  & 7 & 2.0 & 50 & 40 & 58 & 1200 & 7.06 & 10.2 & 0.83 & 12 & This work \\ \hline
TEMPO & \textsuperscript{1}H  &  & 7 & 3.9 & 50 & 24 & 27 & 93 & 10.8 & 26.2 & 10.8 & 2.4 & This work \\ 
TEMPO & \textsuperscript{1}H  &  & 7 & 3.3 & 50 & 25 & 29 & 149 & 9.75 & 24.7 & 6.71 & 3.7 & This work \\ 
TEMPO & \textsuperscript{1}H  &  & 7 & 2.0 & 50 & 30 & 32 & 870 & 9.61 & 21.6 & 1.15 & 19 & This work \\ 
\end{tabular}
\end{adjustbox}
\end{center}
\textsuperscript{\emph{a}} The ratio $k_\mathrm{R}^\mathrm{bup}/k_\mathrm{R}^\mathrm{decay}=1$ indicates the absence of relaxation enhancement by MW irradiation.
\textsuperscript{\emph{b}} The sample formulation was broken down to the radical together with its target spins and possible gadolinium (Gd) doping, although the exact sample formulations vary between the different works. The first three samples from this work refer to the natural abundance sample and the second three to the partially deuterated (DNP juice) sample (cf. Figs.~\ref{fig:BupDecayOverview},\ref{fig:Experiment_1compartment_Relaxation_Microwave}). RF correction was applied by the authors to the decay of Ref.~\cite{macholl_trityl_2010} but not to any of the other reported measurements according to our knowledge. The TEMPO data set from Ref. \cite{Jahnig2019} (same sample formulation as for our natural abundance samples apart from the [\textsuperscript{13}C]urea) was recorded in our lab previously although with an older MW set-up with an approximately eight times lower MW power at the sample (see Methods), possibly explaining the discrepancy to the measurements reported herein.
\end{table*}

Trityl-based dissolution DNP (dDNP) of pyruvic acid is one of the most important medical applications of dissolution DNP while TEMPO and other related aminoxyl-radicals are typically used in NMR applications of DNP. 
Thus, we focus on these materials in the following discussion. 
The available data in the literature is summarized in Tab.~\ref{Tab:RelaxationEnhancementLiterature}.
To be able to analyze the data, the following information was extracted from the literature: (i) temperature and static magnetic field to calculate the electron polarization, (ii) absolute steady-state polarization, (iii) build-up ($\tau_\mathrm{bup}$) and (iv) decay-time ($\tau_\mathrm{decay}$) constants. 

The relaxation-rate constant extracted from the build-up $k_\mathrm{R}^\mathrm{bup}$ (compare Eqs.~\ref{eq:1compartment_bup_solution}) and decay $k_\mathrm{R}^\mathrm{decay}$ are compared to estimate the relaxation enhancement by MW irradiation with a ratio of $k_\mathrm{R}^\mathrm{bup}/k_\mathrm{R}^\mathrm{decay}=1$ indicating no relaxation enhancement in the sample. 
The estimated ratios for the relaxation enhancements are sensitive to inaccuracies in the thermal equilibrium measurement as this affects the steady-state polarization quoted.
The perturbing effects of monitoring RF pulses can be minimized by using few and small flip-angle pulses or correcting for the pulses\cite{von_witte_modelling_2023}.

In the following, relaxation enhancement by MW irradiation depending on the experimental conditions (temperature and field) as well as sample formulation (radical, concentration and target spins) are discussed.
Trends are first discussed for trityl-based and then for TEMPO-based samples.

For trityl radicals and direct DNP to \textsuperscript{13}C nuclei, the relaxation enhancement is more pronounced at lower temperatures and lower fields.
At 3.35\;T, the relaxation enhancement is nearly absent at 3.5\;K, grows to more than fourfold at 1.3\;K and sixfold at 1.0\;K. 
Increasing the field strength reduces the relaxation enhancement in the range of 3.35 and 10.1\;T and temperatures of 1.0-1.3\;K.
Higher trityl (electron) concentrations lead to pronounced increases of the relaxation enhancement.
For example, at 10.1\;T and 1.3\;K, relaxation enhancement is absent for a 15\;mM sample while a twelvefold relaxation enhancement is found for 45\;mM of trityl - resulting in twelvefold shortening of $\tau_\mathrm{bup}$ while preserving steady-state polarization $P_0$. 
The relaxation enhancement at higher electron concentrations can be reduced by sample doping with gadolinium (Gd).

Raising the temperature to 20\;K (and DNP to the \textsuperscript{1}H nuclei), the relaxation enhancement is nearly completely suppressed or masked by increased thermal relaxation even at very high trityl concentrations of 100\;mM.

For TEMPO with DNP performed to \textsuperscript{1}H or \textsuperscript{13}C and 4.2\;K, no field dependence is evident between measurement at 3.35 and 7\;T.
The elevated temperature complicates discrimination between temperature-related and radical-related (DNP mechanism) changes.
The relaxation enhancement appears slightly larger for \textsuperscript{1}H compared to \textsuperscript{13}C although it is not increasing beyond a factor of 2.2.

%As already discussed in Fig.~\ref{fig:BupDecayOverview}c, a pronounced temperature dependence was observed in this work for TEMPO and \textsuperscript{1}H DNP with the temperature dependence resulting from the MW-off relaxation rates.

Two general trends may be deduced from Tab.~\ref{Tab:RelaxationEnhancementLiterature}:
\begin{enumerate}[wide, label=(\roman*), labelwidth=!, labelindent=0pt]
    \item The relaxation enhancement by MW irradiation requires elevated electron concentrations and electron dipolar couplings causing electron spectral diffusion (eSD). Thus, suppression of eSD should reduce the relaxation enhancement. This is in agreement with the results of Gd doping, shortening the electronic relaxation times \cite{Ardenkjaer-Larsen2008,Lumata2013}, or higher magnetic fields, broadening the EPR line and shortening the electronic relaxation \cite{Lumata2013}.
    \item Lower temperatures increase the relaxation enhancement. Lower temperatures prolong the electronic relaxation times and increase the electron polarization. Together, these determine the nuclear relaxation times near liquid helium temperature as all nuclear relaxation can be assumed to originate from paramagnetic relaxation. With lower thermal relaxation, the relaxation enhancement by MW irradiation contributes increasingly more to the total relaxation.
\end{enumerate}

In the following we will discuss possible explanations for the relaxation enhancement by MW irradiation.
MW irradiation saturates parts of the electron line, creating a polarization difference between different parts of the electron line. 
At liquid-He temperatures, nuclear relaxation is most likely due to electron flip-flops \cite{Wenckebach}.
Cross effect and thermal mixing DNP convert this polarization difference into nuclear hyperpolarization.
Assuming that the non-irradiated part of the electron line remains at the thermal electron polarization ($A=P_e^\mathrm{thermal}$) while the irradiated part of the electron line is saturated to $P_e^\mathrm{MW}$, nuclear relaxation for electrons with different polarization is proportional to $1-P_e^\mathrm{thermal}P_e^\mathrm{MW}$ \cite{AbragamGoldman,Wenckebach} (a similar expression is derived for direct nuclear (paramagnetic) relaxation in Ref.~\cite{Wenckebach}).
Hence, nuclear relaxation and DNP injection scale with the electron saturation, leading to a MW power independent steady-state polarization (for small enough thermal relaxation, cf. Fig.~\ref{fig:Experiment_1compartment_Relaxation_Microwave}d).
For MW irradiation at the center of the electron line, the higher electron spectral density and relative ease to fulfill the matching condition for triple spin flips results in a frequency dependence with the higher nuclear relaxation at the center of the electron line.
Increased electron spectral diffusion increases the probability for electron flip-flops which cause nuclear relaxation.

Additionally, sample heating due to MW absorption would appear (nearly) frequency independent over the narrow experimental frequency regime \cite{Warren2008}.
A higher sample temperature reduces the thermal electron polarization which increases the relaxation scaling as $1-(P_e^\mathrm{thermal})^2$.
This might explain the slightly decreasing steady-state polarization for the highest MW powers (see Fig.~\ref{fig:Experiment_1compartment_Relaxation_Microwave}d) and enhanced relaxation for far off-resonant MW irradiation (see Fig.~\ref{fig:Experiment_1compartment_Relaxation_Microwave}a) compared to the MW off situation.
For elevated temperature, the thermal relaxation following a $1-(P_e^\mathrm{thermal})^2$ scaling is large such that a relaxation enhancement by MW irradiation is difficult to observe.

\section*{Conclusions}

Microwave (MW) irradiation in DNP generates not only hyperpolarization of the nuclei but can cause a relaxation enhancement of the nuclei by more than a factor of ten at temperatures of a few Kelvin compared to the decay of nuclear polarization without MW irradiation. 
This is similar to the experimentally observed shortened relaxation times $T_{1\rho}$ \cite{Bornet2016} and $T_2'$ \cite{Guarin2022} under MW irradiation at liquid-helium temperatures.
The relaxation enhancement was measured on \textsuperscript{1}H in samples consisting of a glassy matrix containing TEMPO radicals. 
A literature survey shows that the phenomenon is more general and can also be observed in trityl-based samples, suggesting a radical independent origin. 
The largest relaxation enhancement was obtained with the microwave (MW) irradiation on the EPR resonance frequency, suggesting a resonant relaxation mechanism beyond possible sample heating.
% Both the relaxation enhancement by MW irradiation as well as the build-up time constant follow a square root behavior, resulting in a nearly MW power independent steady-state polarization at higher MW powers.
% The detailed origin of the resonant relaxation enhancement requires further studies but we suspect that it originates from the partial saturation of the electrons increasing the probability of electron flip-flop processes that lead to stronger paramagnetic relaxation.
% If the DNP injection could be kept constant while suppressing the relaxation enhancement by MW irradiation, nuclear polarization enhancements would be boosted significantly. 
% Partially this can be achieved by Gd doping that seems to reduce the influence of MW irradiation on the decay of nuclear polarization.

\section*{Author Contributions}
The research was conceptualized by GvW, SK and ME. 
GvW performed the experiments with help from AH. 
GvW analyzed the data and prepared the original draft. 
SK and ME acquired funding, provided resources and supervision. 
All authors reviewed and edited the draft.

\section*{Conflicts of interest}
The authors declare that they have no competing interests.

\section*{Acknowledgements}
We thank Gian-Marco Camenisch for help with sample preparation and  Jan Henrik Ardenkjær-Larsen for additional data beyond the published \cite{Ardenkjaer-Larsen2019}. GvW thanks Konstantin Tamarov for insightful discussions.

ME acknowledges support by the Schweizerischer Nationalfonds zur Förderung der Wissenschaftlichen Forschung (grant no. 200020\_188988 and 200020\_219375).
Financial support of the Horizon 2020 FETFLAG MetaboliQs grant is gratefully acknowledged.

%\section*{Materials \& Correspondence}
%All data and MATLAB scripts can be found under DOI:10.3929/ethz-b-000606640. Further correspondence should be addressed to Matthias Ernst (maer@ethz.ch).

%%%END OF MAIN TEXT%%%


\balance

\renewcommand\refname{Notes and references}

%%%REFERENCES%%%
\bibliography{library,libraryZotero,notes} %You need to replace "rsc" on this line with the name of your .bib file

\providecommand*{\mcitethebibliography}{\thebibliography}
\csname @ifundefined\endcsname{endmcitethebibliography}
{\let\endmcitethebibliography\endthebibliography}{}
\begin{mcitethebibliography}{37}
\providecommand*{\natexlab}[1]{#1}
\providecommand*{\mciteSetBstSublistMode}[1]{}
\providecommand*{\mciteSetBstMaxWidthForm}[2]{}
\providecommand*{\mciteBstWouldAddEndPuncttrue}
  {\def\EndOfBibitem{\unskip.}}
\providecommand*{\mciteBstWouldAddEndPunctfalse}
  {\let\EndOfBibitem\relax}
\providecommand*{\mciteSetBstMidEndSepPunct}[3]{}
\providecommand*{\mciteSetBstSublistLabelBeginEnd}[3]{}
\providecommand*{\EndOfBibitem}{}
\mciteSetBstSublistMode{f}
\mciteSetBstMaxWidthForm{subitem}
{(\emph{\alph{mcitesubitemcount}})}
\mciteSetBstSublistLabelBeginEnd{\mcitemaxwidthsubitemform\space}
{\relax}{\relax}

\bibitem[Ardenkjaer-Larsen \emph{et~al.}(2015)Ardenkjaer-Larsen, Boebinger,
  Comment, Duckett, Edison, Engelke, Griesinger, Griffin, Hilty, Maeda, Parigi,
  Prisner, Ravera, Van~Bentum, Vega, Webb, Luchinat, Schwalbe, and
  Frydman]{ardenkjaer-larsen_facing_2015}
J.~H. Ardenkjaer-Larsen, G.~S. Boebinger, A.~Comment, S.~Duckett, A.~S. Edison,
  F.~Engelke, C.~Griesinger, R.~G. Griffin, C.~Hilty, H.~Maeda, G.~Parigi,
  T.~Prisner, E.~Ravera, J.~Van~Bentum, S.~Vega, A.~Webb, C.~Luchinat,
  H.~Schwalbe and L.~Frydman, \emph{Angewandte Chemie - International Edition},
  2015, \textbf{54}, 9162--9185\relax
\mciteBstWouldAddEndPuncttrue
\mciteSetBstMidEndSepPunct{\mcitedefaultmidpunct}
{\mcitedefaultendpunct}{\mcitedefaultseppunct}\relax
\EndOfBibitem
\bibitem[{Lilly Thankamony} \emph{et~al.}(2017){Lilly Thankamony}, Wittmann,
  Kaushik, and Corzilius]{Thankamony2017}
A.~S. {Lilly Thankamony}, J.~J. Wittmann, M.~Kaushik and B.~Corzilius,
  \emph{Progress in Nuclear Magnetic Resonance Spectroscopy}, 2017,
  \textbf{102-103}, 120--195\relax
\mciteBstWouldAddEndPuncttrue
\mciteSetBstMidEndSepPunct{\mcitedefaultmidpunct}
{\mcitedefaultendpunct}{\mcitedefaultseppunct}\relax
\EndOfBibitem
\bibitem[Nelson \emph{et~al.}(2013)Nelson, Kurhanewicz, Vigneron, Larson,
  Harzstark, Ferrone, {Van Criekinge}, Chang, Bok, Park, Reed, Carvajal, Small,
  Munster, Weinberg, Ardenkjaer-Larsen, Chen, Hurd, Odegardstuen, Robb, Tropp,
  and Murray]{Nelson2013}
S.~J. Nelson, J.~Kurhanewicz, D.~B. Vigneron, P.~E. Larson, A.~L. Harzstark,
  M.~Ferrone, M.~{Van Criekinge}, J.~W. Chang, R.~Bok, I.~Park, G.~Reed,
  L.~Carvajal, E.~J. Small, P.~Munster, V.~K. Weinberg, J.~H.
  Ardenkjaer-Larsen, A.~P. Chen, R.~E. Hurd, L.~I. Odegardstuen, F.~J. Robb,
  J.~Tropp and J.~A. Murray, \emph{Science Translational Medicine}, 2013,
  \textbf{5}, 198ra108\relax
\mciteBstWouldAddEndPuncttrue
\mciteSetBstMidEndSepPunct{\mcitedefaultmidpunct}
{\mcitedefaultendpunct}{\mcitedefaultseppunct}\relax
\EndOfBibitem
\bibitem[Mugler and Altes(2013)]{Mugler2013}
J.~P. Mugler and T.~A. Altes, \emph{Journal of Magnetic Resonance Imaging},
  2013, \textbf{37}, 313--331\relax
\mciteBstWouldAddEndPuncttrue
\mciteSetBstMidEndSepPunct{\mcitedefaultmidpunct}
{\mcitedefaultendpunct}{\mcitedefaultseppunct}\relax
\EndOfBibitem
\bibitem[Wang \emph{et~al.}(2019)Wang, Ohliger, Larson, Gordon, Bok, Slater,
  Villanueva-Meyer, Hess, Kurhanewicz, and Vigneron]{Wang2019a}
Z.~J. Wang, M.~A. Ohliger, P.~E. Larson, J.~W. Gordon, R.~A. Bok, J.~Slater,
  J.~E. Villanueva-Meyer, C.~P. Hess, J.~Kurhanewicz and D.~B. Vigneron,
  \emph{Radiology}, 2019, \textbf{291}, 273--284\relax
\mciteBstWouldAddEndPuncttrue
\mciteSetBstMidEndSepPunct{\mcitedefaultmidpunct}
{\mcitedefaultendpunct}{\mcitedefaultseppunct}\relax
\EndOfBibitem
\bibitem[Gallagher \emph{et~al.}(2020)Gallagher, Woitek, McLean, Gill, Garcia,
  Provenzano, Riemer, Kaggie, Chhabra, Ursprung, Grist, Daniels, Zaccagna,
  Laurent, Locke, Hilborne, Frary, Torheim, Boursnell, Schiller, Patterson,
  Slough, Carmo, Kane, Biggs, Harrison, Deen, Patterson, Lanz, Kingsbury, Ross,
  Basu, Baird, Lomas, Sala, Wason, Rueda, Chin, Wilkinson, Graves, Abraham,
  Gilbert, Caldas, and Brindle]{Gallagher2020}
F.~A. Gallagher, R.~Woitek, M.~A. McLean, A.~B. Gill, R.~M. Garcia,
  E.~Provenzano, F.~Riemer, J.~Kaggie, A.~Chhabra, S.~Ursprung, J.~T. Grist,
  C.~J. Daniels, F.~Zaccagna, M.~C. Laurent, M.~Locke, S.~Hilborne, A.~Frary,
  T.~Torheim, C.~Boursnell, A.~Schiller, I.~Patterson, R.~Slough, B.~Carmo,
  J.~Kane, H.~Biggs, E.~Harrison, S.~S. Deen, A.~Patterson, T.~Lanz,
  Z.~Kingsbury, M.~Ross, B.~Basu, R.~Baird, D.~J. Lomas, E.~Sala, J.~Wason,
  O.~M. Rueda, S.~F. Chin, I.~B. Wilkinson, M.~J. Graves, J.~E. Abraham, F.~J.
  Gilbert, C.~Caldas and K.~M. Brindle, \emph{Proceedings of the National
  Academy of Sciences of the United States of America}, 2020, \textbf{117},
  2092--2098\relax
\mciteBstWouldAddEndPuncttrue
\mciteSetBstMidEndSepPunct{\mcitedefaultmidpunct}
{\mcitedefaultendpunct}{\mcitedefaultseppunct}\relax
\EndOfBibitem
\bibitem[Fuetterer \emph{et~al.}(2022)Fuetterer, Traechtler, Busch, Peereboom,
  Dounas, Manka, Weisskopf, Cesarovic, Stoeck, and Kozerke]{Fuetterer2022}
M.~Fuetterer, J.~Traechtler, J.~Busch, S.~M. Peereboom, A.~Dounas, R.~Manka,
  M.~Weisskopf, N.~Cesarovic, C.~T. Stoeck and S.~Kozerke, \emph{JACC:
  Cardiovascular Imaging}, 2022, \textbf{15}, 2051--2064\relax
\mciteBstWouldAddEndPuncttrue
\mciteSetBstMidEndSepPunct{\mcitedefaultmidpunct}
{\mcitedefaultendpunct}{\mcitedefaultseppunct}\relax
\EndOfBibitem
\bibitem[Crabb and Meyer(1997)]{Crabb1997}
D.~G. Crabb and W.~Meyer, \emph{Annual Review of Nuclear and Particle Science},
  1997, \textbf{47}, 67--109\relax
\mciteBstWouldAddEndPuncttrue
\mciteSetBstMidEndSepPunct{\mcitedefaultmidpunct}
{\mcitedefaultendpunct}{\mcitedefaultseppunct}\relax
\EndOfBibitem
\bibitem[Kowalska and Neyens(2021)]{Kowalska2021}
M.~Kowalska and G.~Neyens, \emph{Nuclear Physics News}, 2021, \textbf{31},
  14--18\relax
\mciteBstWouldAddEndPuncttrue
\mciteSetBstMidEndSepPunct{\mcitedefaultmidpunct}
{\mcitedefaultendpunct}{\mcitedefaultseppunct}\relax
\EndOfBibitem
\bibitem[Ajoy \emph{et~al.}(2019)Ajoy, Safvati, Nazaryan, Oon, Han, Raghavan,
  Nirodi, Aguilar, Liu, Cai, Lv, Druga, Ramanathan, Reimer, Meriles, Suter, and
  Pines]{Ajoy2019}
A.~Ajoy, B.~Safvati, R.~Nazaryan, J.~T. Oon, B.~Han, P.~Raghavan, R.~Nirodi,
  A.~Aguilar, K.~Liu, X.~Cai, X.~Lv, E.~Druga, C.~Ramanathan, J.~A. Reimer,
  C.~A. Meriles, D.~Suter and A.~Pines, \emph{Nature Communications}, 2019,
  \textbf{10}, 1--12\relax
\mciteBstWouldAddEndPuncttrue
\mciteSetBstMidEndSepPunct{\mcitedefaultmidpunct}
{\mcitedefaultendpunct}{\mcitedefaultseppunct}\relax
\EndOfBibitem
\bibitem[Bucher \emph{et~al.}(2020)Bucher, Glenn, Park, Lukin, and
  Walsworth]{Bucher2020}
D.~B. Bucher, D.~R. Glenn, H.~Park, M.~D. Lukin and R.~L. Walsworth,
  \emph{Physical Review X}, 2020, \textbf{10}, 21053\relax
\mciteBstWouldAddEndPuncttrue
\mciteSetBstMidEndSepPunct{\mcitedefaultmidpunct}
{\mcitedefaultendpunct}{\mcitedefaultseppunct}\relax
\EndOfBibitem
\bibitem[{\'{A}}lvarez \emph{et~al.}(2015){\'{A}}lvarez, Bretschneider,
  Fischer, London, Kanda, Onoda, Isoya, Gershoni, and Frydman]{Alvarez2015}
G.~A. {\'{A}}lvarez, C.~O. Bretschneider, R.~Fischer, P.~London, H.~Kanda,
  S.~Onoda, J.~Isoya, D.~Gershoni and L.~Frydman, \emph{Nature Communications},
  2015, \textbf{6}, 1--2\relax
\mciteBstWouldAddEndPuncttrue
\mciteSetBstMidEndSepPunct{\mcitedefaultmidpunct}
{\mcitedefaultendpunct}{\mcitedefaultseppunct}\relax
\EndOfBibitem
\bibitem[Broadway \emph{et~al.}(2018)Broadway, Tetienne, Stacey, Wood, Simpson,
  Hall, and Hollenberg]{Broadway2018}
D.~A. Broadway, J.~P. Tetienne, A.~Stacey, J.~D. Wood, D.~A. Simpson, L.~T.
  Hall and L.~C. Hollenberg, \emph{Nature Communications}, 2018, \textbf{9},
  1--8\relax
\mciteBstWouldAddEndPuncttrue
\mciteSetBstMidEndSepPunct{\mcitedefaultmidpunct}
{\mcitedefaultendpunct}{\mcitedefaultseppunct}\relax
\EndOfBibitem
\bibitem[McCamey \emph{et~al.}(2009)McCamey, {Van Tol}, Morley, and
  Boehme]{McCamey2009}
D.~R. McCamey, J.~{Van Tol}, G.~W. Morley and C.~Boehme, \emph{Physical Review
  Letters}, 2009, \textbf{102}, 1--4\relax
\mciteBstWouldAddEndPuncttrue
\mciteSetBstMidEndSepPunct{\mcitedefaultmidpunct}
{\mcitedefaultendpunct}{\mcitedefaultseppunct}\relax
\EndOfBibitem
\bibitem[Gangloff \emph{et~al.}(2019)Gangloff, {\'{E}}thier-Majcher, Lang,
  Denning, Bodey, Jackson, Clarke, Hugues, {Le Gall}, and
  Atat{\"{u}}re]{Gangloff2019}
D.~A. Gangloff, G.~{\'{E}}thier-Majcher, C.~Lang, E.~V. Denning, J.~H. Bodey,
  D.~M. Jackson, E.~Clarke, M.~Hugues, C.~{Le Gall} and M.~Atat{\"{u}}re,
  \emph{Science}, 2019, \textbf{364}, 62--66\relax
\mciteBstWouldAddEndPuncttrue
\mciteSetBstMidEndSepPunct{\mcitedefaultmidpunct}
{\mcitedefaultendpunct}{\mcitedefaultseppunct}\relax
\EndOfBibitem
\bibitem[Eills \emph{et~al.}(2023)Eills, Budker, Cavagnero, Chekmenev, Elliott,
  Jannin, Lesage, Matysik, Meersmann, Prisner, Reimer, Yang, and
  Koptyug]{Eills2023}
J.~Eills, D.~Budker, S.~Cavagnero, E.~Y. Chekmenev, S.~J. Elliott, S.~Jannin,
  A.~Lesage, J.~Matysik, T.~Meersmann, T.~Prisner, J.~A. Reimer, H.~Yang and
  I.~V. Koptyug, \emph{Chemical Reviews}, 2023, \textbf{123}, 1417--1551\relax
\mciteBstWouldAddEndPuncttrue
\mciteSetBstMidEndSepPunct{\mcitedefaultmidpunct}
{\mcitedefaultendpunct}{\mcitedefaultseppunct}\relax
\EndOfBibitem
\bibitem[von Witte \emph{et~al.}(2023)von Witte, Ernst, and
  Kozerke]{von_witte_modelling_2023}
G.~von Witte, M.~Ernst and S.~Kozerke, \emph{Magnetic Resonance}, 2023,
  \textbf{4}, 175--186\relax
\mciteBstWouldAddEndPuncttrue
\mciteSetBstMidEndSepPunct{\mcitedefaultmidpunct}
{\mcitedefaultendpunct}{\mcitedefaultseppunct}\relax
\EndOfBibitem
\bibitem[Bornet \emph{et~al.}(2016)Bornet, Pinon, Jhajharia, Baudin, Ji,
  Emsley, Bodenhausen, Ardenkjaer-Larsen, and Jannin]{Bornet2016}
A.~Bornet, A.~Pinon, A.~Jhajharia, M.~Baudin, X.~Ji, L.~Emsley, G.~Bodenhausen,
  J.~H. Ardenkjaer-Larsen and S.~Jannin, \emph{Physical Chemistry Chemical
  Physics}, 2016, \textbf{18}, 30530--30535\relax
\mciteBstWouldAddEndPuncttrue
\mciteSetBstMidEndSepPunct{\mcitedefaultmidpunct}
{\mcitedefaultendpunct}{\mcitedefaultseppunct}\relax
\EndOfBibitem
\bibitem[Guarin \emph{et~al.}(2022)Guarin, Carnevale, Baudin, Pelupessy,
  Abergel, and Bodenhausen]{Guarin2022}
D.~Guarin, D.~Carnevale, M.~Baudin, P.~Pelupessy, D.~Abergel and
  G.~Bodenhausen, \emph{Journal of Physical Chemistry Letters}, 2022,
  \textbf{13}, 175--182\relax
\mciteBstWouldAddEndPuncttrue
\mciteSetBstMidEndSepPunct{\mcitedefaultmidpunct}
{\mcitedefaultendpunct}{\mcitedefaultseppunct}\relax
\EndOfBibitem
\bibitem[J{\"{a}}hnig \emph{et~al.}(2017)J{\"{a}}hnig, Kwiatkowski, D{\"{a}}pp,
  Hunkeler, Meier, Kozerke, and Ernst]{Jahnig2017}
F.~J{\"{a}}hnig, G.~Kwiatkowski, A.~D{\"{a}}pp, A.~Hunkeler, B.~H. Meier,
  S.~Kozerke and M.~Ernst, \emph{Physical Chemistry Chemical Physics}, 2017,
  \textbf{19}, 19196--19204\relax
\mciteBstWouldAddEndPuncttrue
\mciteSetBstMidEndSepPunct{\mcitedefaultmidpunct}
{\mcitedefaultendpunct}{\mcitedefaultseppunct}\relax
\EndOfBibitem
\bibitem[J{\"{a}}hnig \emph{et~al.}(2019)J{\"{a}}hnig, Himmler, Kwiatkowski,
  D{\"{a}}pp, Hunkeler, Kozerke, and Ernst]{Jahnig2019}
F.~J{\"{a}}hnig, A.~Himmler, G.~Kwiatkowski, A.~D{\"{a}}pp, A.~Hunkeler,
  S.~Kozerke and M.~Ernst, \emph{Journal of Magnetic Resonance}, 2019,
  \textbf{303}, 91--104\relax
\mciteBstWouldAddEndPuncttrue
\mciteSetBstMidEndSepPunct{\mcitedefaultmidpunct}
{\mcitedefaultendpunct}{\mcitedefaultseppunct}\relax
\EndOfBibitem
\bibitem[Himmler \emph{et~al.}(2022)Himmler, Albannay, {Von Witte}, Kozerke,
  and Ernst]{Himmler2022}
A.~Himmler, M.~M. Albannay, G.~{Von Witte}, S.~Kozerke and M.~Ernst,
  \emph{Magnetic Resonance}, 2022, \textbf{3}, 203--209\relax
\mciteBstWouldAddEndPuncttrue
\mciteSetBstMidEndSepPunct{\mcitedefaultmidpunct}
{\mcitedefaultendpunct}{\mcitedefaultseppunct}\relax
\EndOfBibitem
\bibitem[Saliba \emph{et~al.}(2017)Saliba, Sesti, Scott, Albert, Choi, Alaniva,
  Gao, and Barnes]{Saliba2017}
E.~P. Saliba, E.~L. Sesti, F.~J. Scott, B.~J. Albert, E.~J. Choi, N.~Alaniva,
  C.~Gao and A.~B. Barnes, \emph{Journal of the American Chemical Society},
  2017, \textbf{139}, 6310--6313\relax
\mciteBstWouldAddEndPuncttrue
\mciteSetBstMidEndSepPunct{\mcitedefaultmidpunct}
{\mcitedefaultendpunct}{\mcitedefaultseppunct}\relax
\EndOfBibitem
\bibitem[Tan and Griffin(2022)]{Tan2022}
K.~O. Tan and R.~G. Griffin, \emph{Journal of Chemical Physics}, 2022,
  \textbf{156}, 1--6\relax
\mciteBstWouldAddEndPuncttrue
\mciteSetBstMidEndSepPunct{\mcitedefaultmidpunct}
{\mcitedefaultendpunct}{\mcitedefaultseppunct}\relax
\EndOfBibitem
\bibitem[Stern \emph{et~al.}(2021)Stern, Cousin, Mentink-Vigier, Pinon,
  Elliott, Cala, and Jannin]{Stern2021}
Q.~Stern, S.~F. Cousin, F.~Mentink-Vigier, A.~C. Pinon, S.~J. Elliott, O.~Cala
  and S.~Jannin, \emph{Science Advances}, 2021, \textbf{7}, 1--14\relax
\mciteBstWouldAddEndPuncttrue
\mciteSetBstMidEndSepPunct{\mcitedefaultmidpunct}
{\mcitedefaultendpunct}{\mcitedefaultseppunct}\relax
\EndOfBibitem
\bibitem[Leavesley \emph{et~al.}(2018)Leavesley, Jain, Kamniker, Zhang, Rajca,
  Rajca, and Han]{Leavesley2018a}
A.~Leavesley, S.~Jain, I.~Kamniker, H.~Zhang, S.~Rajca, A.~Rajca and S.~Han,
  \emph{Physical Chemistry Chemical Physics}, 2018, \textbf{20},
  27646--27657\relax
\mciteBstWouldAddEndPuncttrue
\mciteSetBstMidEndSepPunct{\mcitedefaultmidpunct}
{\mcitedefaultendpunct}{\mcitedefaultseppunct}\relax
\EndOfBibitem
\bibitem[Prisco \emph{et~al.}(2021)Prisco, Pinon, Emsley, and
  Chmelka]{Prisco2021}
N.~A. Prisco, A.~C. Pinon, L.~Emsley and B.~F. Chmelka, \emph{Physical
  Chemistry Chemical Physics}, 2021, \textbf{23}, 1006--1020\relax
\mciteBstWouldAddEndPuncttrue
\mciteSetBstMidEndSepPunct{\mcitedefaultmidpunct}
{\mcitedefaultendpunct}{\mcitedefaultseppunct}\relax
\EndOfBibitem
\bibitem[Batel(2013)]{BatelThesis}
M.~Batel, \emph{Ph.D. thesis}, ETH Z{\"{u}}rich, 2013\relax
\mciteBstWouldAddEndPuncttrue
\mciteSetBstMidEndSepPunct{\mcitedefaultmidpunct}
{\mcitedefaultendpunct}{\mcitedefaultseppunct}\relax
\EndOfBibitem
\bibitem[Macholl \emph{et~al.}(2010)Macholl, Jóhannesson, and
  Ardenkjaer-Larsen]{macholl_trityl_2010}
S.~Macholl, H.~Jóhannesson and J.~H. Ardenkjaer-Larsen, \emph{Physical
  Chemistry Chemical Physics}, 2010, \textbf{12}, 5804--5817\relax
\mciteBstWouldAddEndPuncttrue
\mciteSetBstMidEndSepPunct{\mcitedefaultmidpunct}
{\mcitedefaultendpunct}{\mcitedefaultseppunct}\relax
\EndOfBibitem
\bibitem[Ardenkj{\ae}r-Larsen \emph{et~al.}(2019)Ardenkj{\ae}r-Larsen, Bowen,
  Petersen, Rybalko, Vinding, Ullisch, and Nielsen]{Ardenkjaer-Larsen2019}
J.~H. Ardenkj{\ae}r-Larsen, S.~Bowen, J.~R. Petersen, O.~Rybalko, M.~S.
  Vinding, M.~Ullisch and N.~C. Nielsen, \emph{Magnetic Resonance in Medicine},
  2019, \textbf{81}, 2184--2194\relax
\mciteBstWouldAddEndPuncttrue
\mciteSetBstMidEndSepPunct{\mcitedefaultmidpunct}
{\mcitedefaultendpunct}{\mcitedefaultseppunct}\relax
\EndOfBibitem
\bibitem[Ard()]{Ardenkjaer-LarsenPrivate}
Private communication with Jan Henrik Ardenkjær-Larsen.\relax
\mciteBstWouldAddEndPunctfalse
\mciteSetBstMidEndSepPunct{\mcitedefaultmidpunct}
{}{\mcitedefaultseppunct}\relax
\EndOfBibitem
\bibitem[Equbal \emph{et~al.}(2020)Equbal, Li, Tabassum, and Han]{Equbal2020}
A.~Equbal, Y.~Li, T.~Tabassum and S.~Han, \emph{Journal of Physical Chemistry
  Letters}, 2020, \textbf{11}, 3718--3723\relax
\mciteBstWouldAddEndPuncttrue
\mciteSetBstMidEndSepPunct{\mcitedefaultmidpunct}
{\mcitedefaultendpunct}{\mcitedefaultseppunct}\relax
\EndOfBibitem
\bibitem[Ardenkjaer-Larsen \emph{et~al.}(2008)Ardenkjaer-Larsen, MacHoll, and
  J{\'{o}}hannesson]{Ardenkjaer-Larsen2008}
J.~H. Ardenkjaer-Larsen, S.~MacHoll and H.~J{\'{o}}hannesson, \emph{Applied
  Magnetic Resonance}, 2008, \textbf{34}, 509--522\relax
\mciteBstWouldAddEndPuncttrue
\mciteSetBstMidEndSepPunct{\mcitedefaultmidpunct}
{\mcitedefaultendpunct}{\mcitedefaultseppunct}\relax
\EndOfBibitem
\bibitem[Lumata \emph{et~al.}(2013)Lumata, Kovacs, Sherry, Malloy, Hill, {Van
  Tol}, Yu, Song, and Merritt]{Lumata2013}
L.~Lumata, Z.~Kovacs, A.~D. Sherry, C.~Malloy, S.~Hill, J.~{Van Tol}, L.~Yu,
  L.~Song and M.~E. Merritt, \emph{Physical Chemistry Chemical Physics}, 2013,
  \textbf{15}, 9800--9807\relax
\mciteBstWouldAddEndPuncttrue
\mciteSetBstMidEndSepPunct{\mcitedefaultmidpunct}
{\mcitedefaultendpunct}{\mcitedefaultseppunct}\relax
\EndOfBibitem
\bibitem[Wenckebach(2016)]{Wenckebach}
T.~Wenckebach, \emph{{Essentials of Dynamic Nuclear Polarization}}, Spindrift,
  2016\relax
\mciteBstWouldAddEndPuncttrue
\mciteSetBstMidEndSepPunct{\mcitedefaultmidpunct}
{\mcitedefaultendpunct}{\mcitedefaultseppunct}\relax
\EndOfBibitem
\bibitem[Abragam and Goldman(1982)]{AbragamGoldman}
A.~Abragam and M.~Goldman, \emph{{Nuclear Magnetism: Order and Disorder}},
  Oxford University Press, 1982\relax
\mciteBstWouldAddEndPuncttrue
\mciteSetBstMidEndSepPunct{\mcitedefaultmidpunct}
{\mcitedefaultendpunct}{\mcitedefaultseppunct}\relax
\EndOfBibitem
\bibitem[Warren and Brandt(2008)]{Warren2008}
S.~G. Warren and R.~E. Brandt, \emph{Journal of Geophysical Research
  Atmospheres}, 2008, \textbf{113}, 1--10\relax
\mciteBstWouldAddEndPuncttrue
\mciteSetBstMidEndSepPunct{\mcitedefaultmidpunct}
{\mcitedefaultendpunct}{\mcitedefaultseppunct}\relax
\EndOfBibitem
\end{mcitethebibliography}


\begin{thebibliography}{1}

\bibitem{Himmler2022}
Aaron Himmler, Mohammed~M Albannay, Gevin {Von Witte}, Sebastian Kozerke, and
  Matthias Ernst.
\newblock {Electroplated waveguides to enhance DNP and EPR spectra of silicon
  and diamond particles}.
\newblock {\em Magnetic Resonance}, 3(2):203--209, 2022.

\bibitem{von_witte_modelling_2023}
Gevin von Witte, Matthias Ernst, and Sebastian Kozerke.
\newblock Modelling and correcting the impact of {RF} pulses for continuous
  monitoring of hyperpolarized {NMR}.
\newblock {\em Magnetic Resonance}, 4:175--186, 2023.

\end{thebibliography}
\bibliographystyle{rsc} %the RSC's .bst file

\end{document}

%----------------------------------------------------------------------------------------
%	SUPPLEMENTARY INFORMATION
%----------------------------------------------------------------------------------------

\onecolumn

% \phantomsection
% \section*{Supplementary information: Relaxation enhancement by MW irradiation limits dynamic nuclear polarization}

%%%%%%%%%% Prefix a "S" to all equations, figures, tables and reset the counter %%%%%%%%%%
\setcounter{equation}{0}
\setcounter{figure}{0}
\setcounter{table}{0}
\setcounter{section}{0}
\setcounter{page}{1}
\makeatletter
\renewcommand{\theequation}{S\arabic{equation}}
\renewcommand{\thefigure}{S\arabic{figure}}
\renewcommand{\thetable}{S\arabic{table}}
\renewcommand{\thesection}{S\arabic{section}}
%\renewcommand{\bibnumfmt}[1]{[S#1]}
%\renewcommand{\citenumfont}[1]{S#1}
%%%%%%%%%% Prefix a "S" to all equations, figures, tables and reset the counter %%%%%%%%%%

\maketitle % Output the title and abstract box

\tableofcontents

\clearpage

\section{Build-up and decay data with and without RF correction} 

\begin{figure*}[!ht]\centering % Using \begin{figure*} makes the figure take up the entire width of the page
	\includegraphics[width=\linewidth]{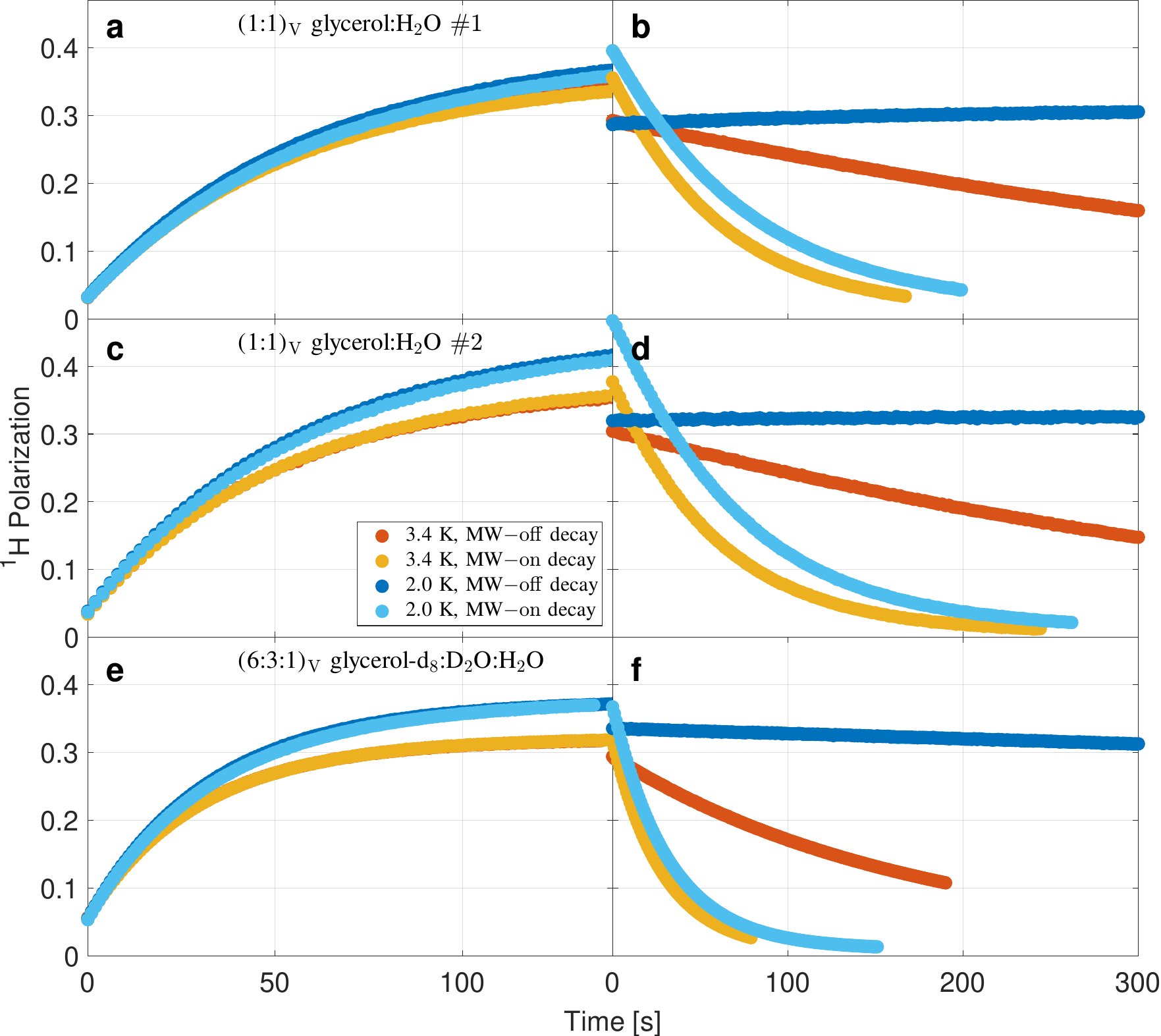}
	\caption{\textbf{Build-up and decay with RF correction} Build-ups \textbf{(a)} and the respective decays \textbf{(b)} at 3.3\;K (red and orange) and 2.0\;K (light and darker blue) with correcting for the effects of radio-frequency (RF) pulses. The MW is set to the same power and frequency (197.05GHz) during all build-ups. During the darker decays (darker blue and red), the MW is switched off. For the lighter decays (3.3\;K in orange and 2.0\;K in light blue), the MW is switched to the central EPR frequency (197.28\;GHz) resulting in zero DNP, as displayed in \ref{fig:BupDecayOverview}d.}
	\label{fig:BupDecayOverview_corr}
\end{figure*}

\begin{figure*}[!ht]\centering % Using \begin{figure*} makes the figure take up the entire width of the page
	\includegraphics[width=\linewidth]{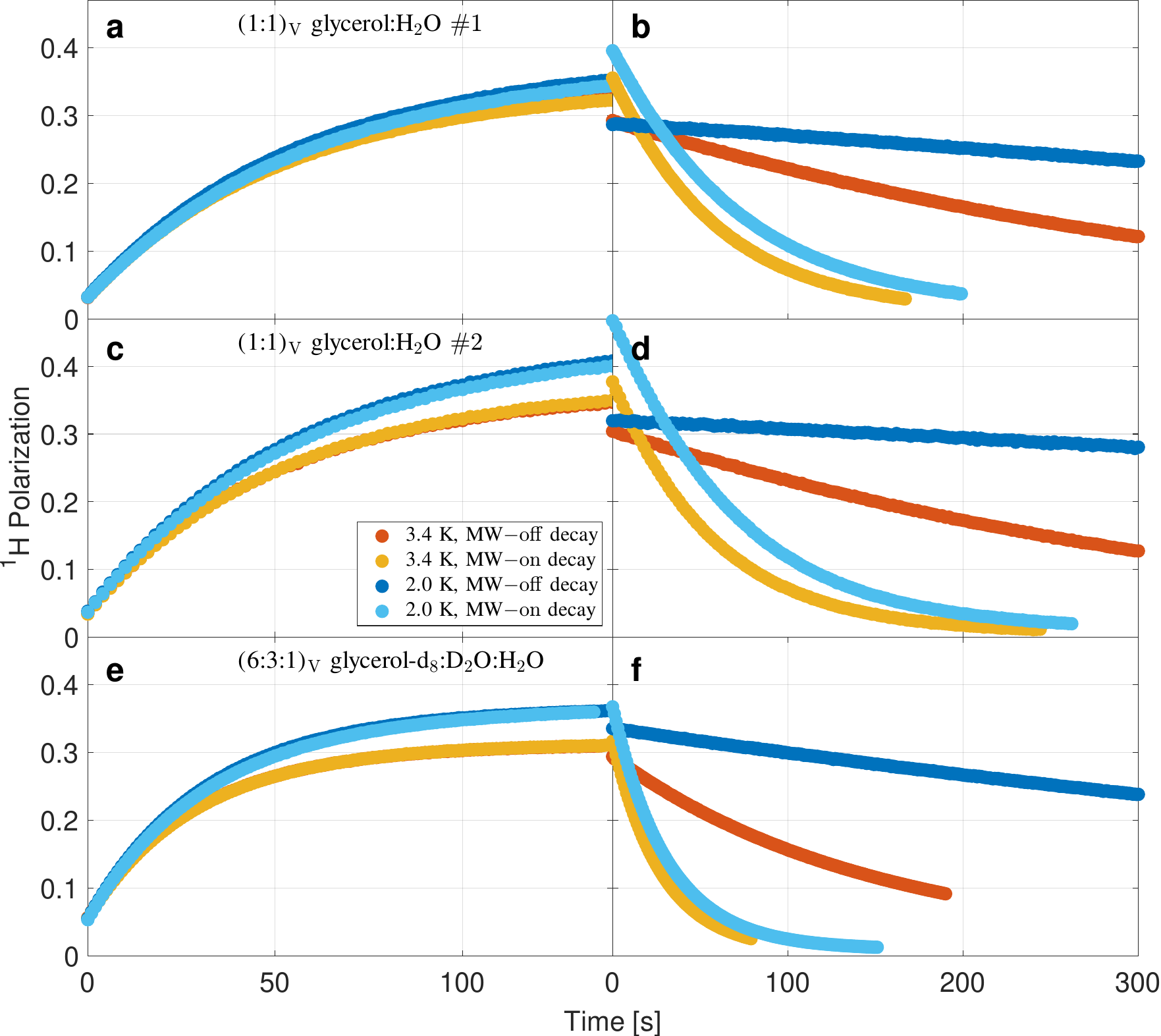}
	\caption{\textbf{Build-up and decay without RF correction} Build-ups \textbf{(a)} and the respective decays \textbf{(b)} at 3.3\;K (red and orange) and 2.0\;K (light and darker blue) without correcting for the effects of radio-frequency (RF) pulses. The MW is set to the same power and frequency (197.05GHz) during all build-ups. During the darker decays (darker blue and red), the MW is switched off. For the lighter decays (3.3\;K in orange and 2.0\;K in light blue), the MW is switched to the central EPR frequency (197.28\;GHz) resulting in zero DNP, as displayed in \ref{fig:BupDecayOverview}d.}
	\label{fig:BupDecayOverview_uncorr}
\end{figure*}

\clearpage

\section{Experimental validation of the one compartment model} \label{sec:SI_ModelValidation}

In order to verify the validity of the extracted model parameters, namely $k_{W}$ and $k_\mathrm{R}$, from our measurements, we performed build-up experiments with amplitude modulation.
In our case, we switched the MW between zero and full output power in a fixed cycle. 
For the choice cycle time, we have two different time scales to consider: (i) It should be much longer than the thermal $T_{1,e}$ (around 500\;$\mu$s under our DNP conditions measured with longitudinal detected (LOD) electron paramagnetic resonance (EPR) \cite{Himmler2022}, probably dominated by electron spectral diffusion) such that we can consider the MW irradiation blocks as continuous wave irradiation and (ii) much shorter than the build-up time such that we can average our modulated irradiation over time and consider it to be a constant irradiation on the time scale of the build-up. 
Based on these consideration, we chose a cycle time of 1\,s. This would even be enough if the LOD EPR-based $T_{1,e}$ is two orders of magnitude shorter than the "true" $T_{1,e}$.
Upon varying the time with full MW output power per cycle, we can translate this into a linear variation of our model parameters, eventually enabling us to disentangle the MW-induced relaxation from the DNP injection. 

\begin{figure*}[!ht]\centering % Using \begin{figure*} makes the figure take up the entire width of the page
	\includegraphics[width=\linewidth]{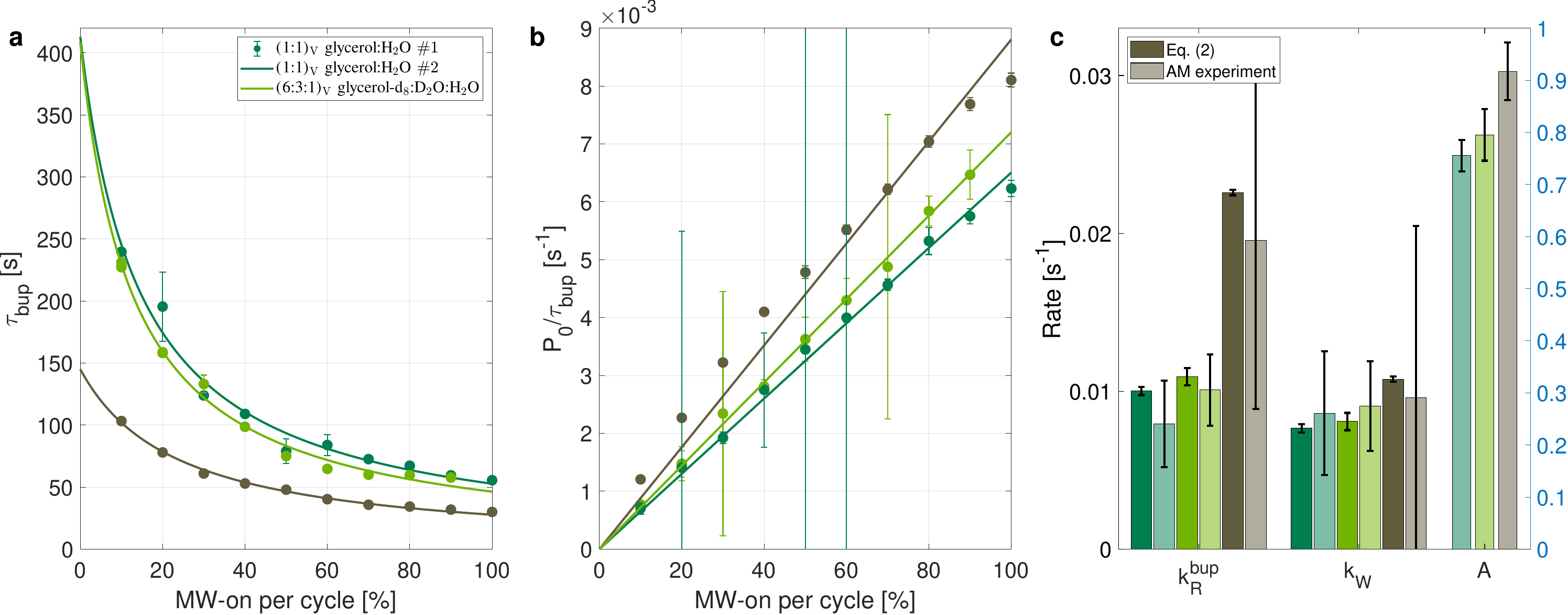}
	\caption{\textbf{Measuring the one compartment model parameters.} \textbf{(a)} We fit the build-up times for different fractions of the MW being switched on to full power (amplitude modulation with full and zero power) with  $(k_{R,\mathrm{res}} + b \cdot x)^{-1}$. $k_{R,\mathrm{res}}$ is a fitted residual relaxation rate (in good agreement with the thermal relaxation rate in MW-off decays, see Fig.~\ref{fig:BupDecayOverview}c or \cite{von_witte_modelling_2023}, $b=k_{W}+k_{R,\mathrm{MW}}$ and $x$ being the MW-on fraction. The green data sets belong to natural abundance and brown-gray (taupe) to the partially deuterated sample as throughout this paper. \textbf{(b)} $P_0/\tau_{\mathrm{bup}}$ shows a linear behaviour as expected from Eq.~\ref{eq:1compartment_bup_P0} with the slope being given by $Ak_{W}$. \textbf{(c)} Comparison of $k_\mathrm{R}^\mathrm{bup}$ and $k_{W}$ through two different analysis approaches: (i, full color) Inserting the measured steady-state polarization and build-up time into Eqs.~\ref{eq:1compartment_bup_solution}, (ii, half transparent) The relaxation rate $k_\mathrm{R}^\mathrm{bup}$ can be estimated from the fit shown in Fig.~\ref{fig:Experiment_1compartment_Relaxation_Microwave}a. With $k_\mathrm{R}^\mathrm{bup}$ at hand, $k_{W}$ can be extracted from the fit shown in part (a) of this figure. Finally, $A$ can be deduced from the linear fit shown in (b).}
	\label{fig:Experiment_1compartment_TheoryParameters}
\end{figure*}

For the amplitude modulated build-ups (MW switched between full and zero power) as shown in Fig.~\ref{fig:Experiment_1compartment_TheoryParameters}, the large error bars are a result of electronics communications problems leading to clear bumps in the data (not shown), visible during processing. 
The affected data sets were cut but often resulted in short, usable build-up durations, leading to relatively large uncertainties in the fit results.

The estimated values of the thermal electron polarization ($A$) appear below the thermal electron polarization based on the experimental conditions (89\% at 3.3\,K and 7\,T) which might be an indication of sample heating from MW absorption as discussed in the main text.

\clearpage

\section{Hypothetical polarization}

\begin{figure}[!ht]\centering % Using \begin{figure*} makes the figure take up the entire width of the page
		\includegraphics[width=0.4\linewidth]{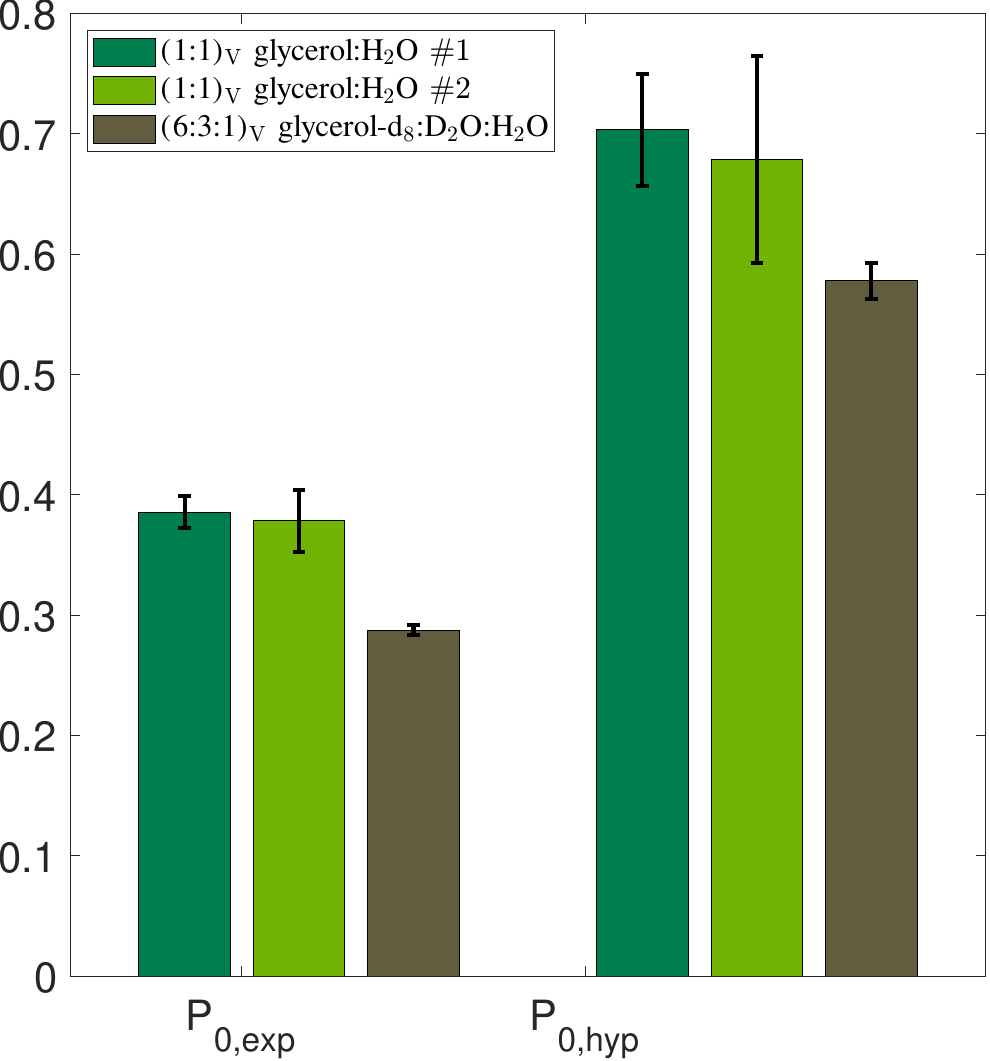}
		\caption{Experimental and hypothetical polarization. The experimentally measured steady-state polarization $P_{0,\mathrm{exp}}$ of the three samples is compared with a hypothetical polarization $P_{0,\mathrm{hyp}}$ without MW-induced relaxation (see thermal relaxation in Fig.~\ref{fig:BupDecayOverview}c). This hypothetical polarization was calculated with Eq.~\ref{eq:1compartment_bup_P0}, the thermal electron polarization $A=0.89$ and $k_{W}$ extracted from the measured build-up time and steady-state polarization (compare Fig.~\ref{fig:Experiment_1compartment_TheoryParameters}c ).}
		\label{fig:HypotheticalPolarization}
	\end{figure}

\section{Measurement of the MW frequency dependence of the relaxation enhancement}

During the measurement it needs to be carefully checked that the chosen MW frequency during the decay does not lead to a significant hyperpolarization as the measured relaxation time in these types of decay is the build-up time. Only for one of the presented data points (196.75\;GHz), a build-up with signals clearly exceeding the thermal signal could be measured with the steady-state polarization being less than ten percent of the used build-up frequency (197.05\;GHz), justifying the usage of the measured decay time as a relaxation rate.

\section{Sample composition}
The faster relaxation of the deuterated sample compared to the natural abundance samples remains unclear. 
The quadrupolar $^2$H spins might either cause (i) relaxation of the \textsuperscript{1}H spins directly, (ii) increase the electron relaxation, (iii) the lower number of spins in the bulk is relaxed faster as each electron (same concentration for all measurements) is able to relax a certain number of spins per unit time or (iv) the lower density of \textsuperscript{1}H spins close to the electron suppresses the transport of polarization from the strongly hyperfine coupled quenched spins into the bulk (larger frequency differences among the nuclear spin pairs involved).
%Our results rather rule out the first option as the bulk relaxation could be set to zero in all our simulations and still achieving an excellent agreement with the experimental results (see Fig.~\ref{fig:Simulation_HypResOn_Power}).

\renewcommand\refname{References}

%%%REFERENCES%%%
\phantomsection
\bibliographystyle{unsrt}
\bibliography{library,libraryZotero,notes}